# Optimized Cryo-CMOS Technology with $V_{TH}$<0.2V and $I_{on}$>1.2mA/μm for High-Peformance Computing

Chang He[1,2,#], Yue Xin[1,2,#], Longfei Yang[3,#], Zewei Wang[1], Zhidong Tang[1,4], Xin Luo[3], Renhe Chen[1], Zirui Wang[1], Shuai Kong[2], Jianli Wang[2], Jianshi Tang[4], Xiaoxu Kang[3], Shoumian Chen[3], Yuhang Zhao[3], Shaojian Hu[3,*], and Xufeng Kou[1,2*]
[1]ShanghaiTech University, Shanghai, China, [2]Zhangjiang Laboratory, Shanghai, China, [3]Shanghai IC Research and Development Center (ICRD), Shanghai, China, [4]Tsinghua University, Beijing, China,
*Email: kouxf@shanghaitech.edu.cn, hushaojian@icrd.com.cn; #Authors contributed equally to this work.

*Abstract*—We report the design-technology co-optimization (DTCO) scheme to develop a 28-nm cryogenic CMOS (Cryo-CMOS) technology for high-performance computing (HPC). The precise adjustment of halo implants manages to compensate the threshold voltage ($V_{TH}$) shift at low temperatures. The optimized NMOS and PMOS transistors, featured by $V_{TH}$<0.2V, sub-threshold swing (*SS*)<30 mV/dec, and on-state current ($I_{on}$)>1.2mA/μm at 77K, warrant a reliable sub-0.6V operation. Moreover, the enhanced driving strength of Cryo-CMOS inherited from a higher transconductance leads to marked improvements in elevating the ring oscillator frequency by 20%, while reducing the power consumption of the compute-intensive cryogenic IC system by 37% at 77K.

## I. Introduction

Cryogenic CMOS has been regarded as a promising HPC platform which could break through the energy and speed boundaries of data processing at room temperature (RT) [1]. Benefiting from the steeper sub-threshold swing, higher channel mobility ($\mu_{ch}$), and diminished leakage current ($I_{off}$), the intrinsic electrical characteristics of the Cryo-CMOS device would behave close to that of an ideal switch [2]. Meanwhile, the lower interconnect resistivity and suppressed thermal noise at cryogenic temperatures help to improve the data transfer rate and signal-to-noise ratio, therefore increasing the bandwidth of the data bus [3]. On the other hand, however, the incomplete ionization of dopants changes the Fermi level position in the bulk well, and the resulting carrier freeze-out effect would lead to a shift of the threshold voltage by $\Delta V_{TH}$ = 0.1~0.3 V (*i.e.*, which depends on the gate geometry) at cryogenic temperatures, as summarized in Fig. 1 [4-6]. In this context, the enlarged $V_{TH}$ inevitably shrinks the overdrive voltage range of the input signal, which not only complicates the circuit design, but also places a roadblock to lowering the supply voltage ($V_{DD}$) in the low-temperature (LT) region [7].

In order to take full advantage of Cryo-CMOS for realizing the performance versus energy benefits, in this work, we report the use of halo implants to tailor the threshold voltage into the sub-0.2V region, and experimentally demonstrate that the optimized 28-nm high-k metal gate (HKMG) Cryo-CMOS technology supports a low-power working mode with $V_{DD}$ = 0.6 V at 77 K. Besides, we show the boosted saturation current and broadened operation range facilitate the design of high-speed cryogenic integrated circuits for HPC applications.

## II. Design-Technology Co-Optimization of Cryo-CMOS by Halo Implantation

In general, the threshold voltage of a transistor is given by
$$V_{TH} = V_{FB} + \phi_S + \sqrt{2q\varepsilon_{Si}N_{ch}\phi_S}/C_{ox} \quad (1)$$
where $V_{FB}$ is the flat-band voltage, $N_{ch}$ is the activated doping level in the channel, $C_{ox}$ is the gate capacitance, and the surface potential at the threshold condition equals to $\phi_S = (2kT/q)\cdot \ln(N_{ch}/n_i)$ [8]. In the 28-nm HKMG technology, it is difficult to modulate the $V_{FB}$ value of PMOS due to the limited choice of suitable gate metal materials; alternatively, we proposed to tune $V_{TH}$ by modifying the halo implant during the CMOS fabrication process (Fig. 2a). From the doping profile obtained by the TCAD simulations in Fig. 2b, it is seen that the localized dopant pocket within the halo structure induces an energy barrier between the bulk and the source/drain regions. As a result, reducing the halo implantation dose would lower the barrier height of this junction, which in turn causes a decrease in the threshold voltage and slightly promotes the current conduction in the inversion layer (Figs. 3a-b). On the contrary, the low halo doping level invariably brings about a weakened control of the transition between the on/off state accompanied by large *SS* and leakage values, yet such issues can be greatly alleviated at cryogenic temperatures (Figs. 3c-d). After optimizing the halo implant condition to make good trade-offs between these electrical parameters, the TCAD-simulated $I_{DS}$-$V_{GS}$ curves of both NMOS and PMOS achieve a nearly zero-voltage turn-on operation at $T$ = 77 K (Fig. 4). Here, it is worth noting that the halo implant has a more pronounced effect in reducing the threshold voltage in wide-channel devices than long-channel ones, because more dopants need to be introduced via ion implantation to affect $N_{ch}$ along the channel width direction (Fig. 5). Guided by the above TCAD results, the key process parameters have been finalized, and the cross-sectional structure of the fabricated Cryo-CMOS device was investigated by transmission electron microscopy (Fig. 6).

## III. Cryo-CMOS Device Characterizations

### A. Temperature-dependent I-V measurements

Temperature-dependent *I-V* measurements were performed on a set of Cryo-CMOS transistors with the same channel length (*L*) of 28 nm but varied width (*W*) from 0.1 μm to 3 μm, and Fig. 7 illustrates the measured $I_{DS}$-$V_{GS}$ transfer characteristics of two optimized NMOS and PMOS devices from $T$ = 10 K to 298 K. Consistent the TCAD simulation results, the adjustment of the doping concentration within the channel results in a small onset gate voltage in reference to the strong-inversion condition, and the sharp switching behavior ensures a negligible leakage current at zero voltage bias in the LT region. Accordingly, by applying the conventional constant current method ($I_{DS} \cdot (L/W) = 10^{-8}$ A), the threshold voltages are found to be 0.109 V (NMOS) and 0.171 V (PMOS) at 77 K. To exclude the influence of self-heating effect at high drain bias, we further adopted the Y-function method to extract the threshold voltage based on [9]

$$I_{DS}/\sqrt{g_{m,lin}} = (\mu_{ch} C_{ox} W V_{DS}/L)^{1/2} \cdot (V_{GS} - V_{TH}) \quad (2)$$

Therefore, by extrapolating the measured $I_{DS}/\sqrt{g_{m,lin}}$ curves in the linear region under a fixed $V_{DS} = 0.05$ V (Fig. 8), the $V_{TH}$ values are found to remain in the sub-0.2V region in the entire temperature region (Fig. 9), thus accomplishing the design goal of the Cryo-CMOS technology.

In view of the driving strength of Cryo-CMOS, the $I_{DS}$-$V_{GS}$ curves of both NMOS and PMOS with the smallest device feature ($W/L = 0.1\mu m/0.03\mu m$) unveil that the channel current increases by 20% when the transistors are cooled down to 10 K (Fig. 10). Likewise, the LT transconductance ($g_m$) under $V_{DS} = 0.9$ V also experiences an evident improvement by 30% than the RT operation, and the maximum value of $g_m(77 K) = 0.25$ mS is observed at $V_{GS} = 0.6$ V (Fig. 11). Along with the scaled-down sub-threshold swing from 105 mV/dec (298K) to 18 mV/dec (10K), the required overdrive voltage ($V_{OV}=V_{GS}-V_{TH}$) to sustain a high on/off ratio decreases monotonically with temperature. For instance, a supplied voltage of 0.6 V is able to meet the $I_{on}/I_{off}=10^7$ baseline at 77 K (Fig. 12), hence validating the high performance of our Cryo-CMOS devices.

### B. Cryo-CMOS device modeling

Temperature-dependent channel mobility values extracted from Eq. (2) display the same evolution trend among all NMOS/PMOS devices: the quenched phonon scattering helps to elevate $\mu_{ch}$ as temperature drops from 298 K, yet it gradually reaches a saturation plateau when $T < 77$ K, mainly due to the presence of the Coulomb scattering (Fig. 13). Such an enhanced low-field channel mobility accelerates the electron transport, and the linear $I_{DSAT}$−$V_{OV}$ curves visualized in Fig. 14 confirm the velocity saturation-induced current saturation in our nano-scale Cryo-CMOS devices. Besides, the zero-bias currents of different transistors follow the universal $I_{off}(T) = I_{off,0} \cdot 10^{T \cdot \eta}$ (i.e., where $\eta$ is a size-dependent fitting parameter) scaling law (Fig. 15), manifesting the effective suppression of the bulk-state leakage path at low temperatures. Afterwards, by taking the aforementioned LT-related mechanisms into consideration, we developed a semi-empirical device compact model to capture the electrical $I_{DS}$-$V_{DS}$ characteristics of the Cryo-CMOS devices under different bias conditions, and the fitting errors are below 6% in the examined temperature region, as shown in Fig. 16.

### C. Comparisions of different Cryo-CMOS technologies

Quantitative comparisons of device performance between three types of transistors (i.e., Cryo-CMOS, uLVT, and RVT) fabricated on the same 28-nm HKMG CMOS process line are presented in Fig. 17. The sub-0.2V $V_{TH}$ realized in the Cryo-CMOS device guarantees a much wider operation range of −0.61V ≤ $V_{OV}$ ≤ +0.84 V at $T$ = 77 K, and the channel current of NMOS (PMOS) at $|V_{OV}|$=0.6V and $|V_{DS}|$=0.9V increases by 19% (36%) and 29% (53%) compared to the uLVT and RVT processes. Correspondingly, the saturated $I_{DSAT}$ reaches 1.6 mA/μm at $V_{DD}$ = 0.9 V, and it stays above the 0.7 mA/μm level even when $V_{DD}$ is reduced to 0.6 V (Fig. 18). In contrast, neither uLVT nor RVT-NMOS can provide sufficient driving strength (i.e., $I_{DSAT}$ < 0.34 mA/μm) under such a low supply voltage. Furthermore, the extended comparison charts among different technology nodes are listed in Figs. 19 and 20 [10]. It is obvious that our optimized Cryo-CMOS devices excel in all categories, where the medium value of $V_{TH}$ decreases by 130~410 mV, and the average $I_{DSAT}$ increases by 0.28~1.26 mA/μm.

## IV. CIRCUIT-LEVEL PERFORMANCE ESTIMATION

To evaluate the speed/power benchmarks of our proposed Cryo-CMOS technology, we subsequently conducted circuit-level simulations on three representative cryogenic circuits, namely a 257-stage ring oscillator (RO), a master-slave-type D-flip flop (DFF), and a digital IC module which implements the advanced encryption standard (AES) algorithm (Fig. 21). In particular, the combination of low-$V_{TH}$ and high-$I_{DSAT}$ empowers a significant enhancement of the RO oscillation frequency over the standard RVT process. As highlighted in Fig. 22a, the Cryo-CMOS-constructed RO circuit can always produce high-quality 200~600 MHz signals, whereas the RVT-based counterpart fails to respond at $V_{DD}$ = 0.6 V when $T$ ≤ 77 K. Concurrently, the propagation delay of DFF also reduces up to 25% when RVT devices are replaced by the Cryo-CMOS ones (Fig. 22b). Finally, considering that the AES algorithm includes numerous shift, matrix multiplication, and XOR operations, it need to consume 2.03 mW to complete a basic AES round operation by RVT-based circuit. Instead, when the optimized low-$V_{TH}$ transistors are incorporated, the total power is reduced to 1.28 mW when the hardware system is operated at 77 K with the same frequency (Fig. 22c).

## V. CONCLUSION

We demonstrated a 28nm Cryo-CMOS technology which exhibits salient device performance in the LT region. The small-$SS$, low-$V_{TH}$, and high-$I_{on}$ merits allow for a low-$V_{DD}$ operation with the on/off ratio up to $10^7$ at 77 K. The enhanced device driving strength enables the high-speed mode of the designed cryogenic circuit. In addition, the optimization strategy elaborated in this work may also extend to other advanced nodes to construct energy-efficient HPC systems.

### ACKNOWLEDGMENT

This work is supported by the National Key R&D Program of China (2023YFB4404000), the NSFC Program (92164104), Zhangjiang Lab Strategic Program, and ShanghaiTech SMDL.

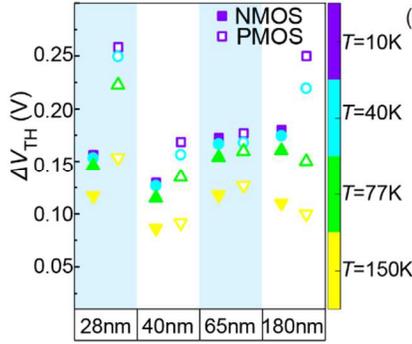

Fig. 1. Carrier freeze-out effect-induced $V_{TH}$ shift in the low-temperature region for different technology nodes.

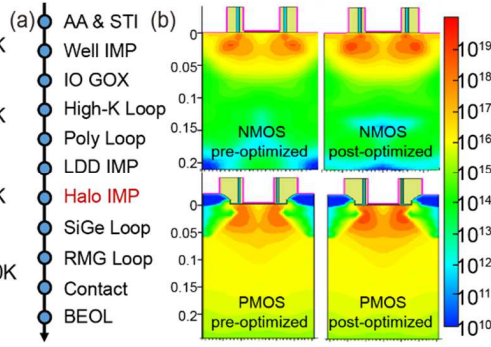

Fig. 2. (a) Process flow of proposed 28nm HKMG Cryo-CMOS technology. (b) TCAD-simulated doping profiles before and after the halo implant adjustments.

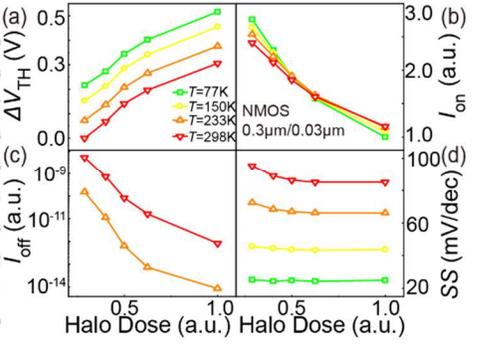

Fig. 3. The evolutions of (a) $\Delta V_{TH}$, (b) $I_{on}$, (c) $I_{off}$, and (d) $SS$ with respect to the halo implantation dose level of NMOS from TCAD simulations.

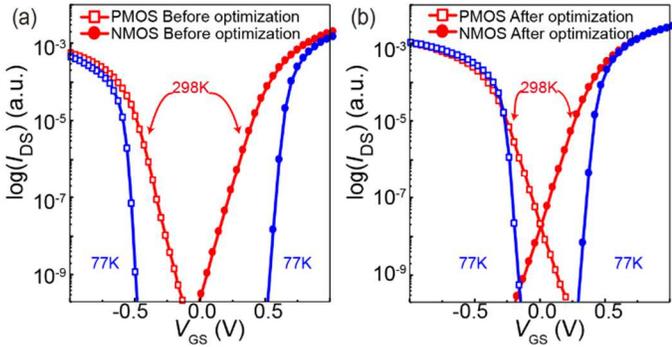

Fig. 4. TCAD simulation results of the $I_{DS}$-$V_{GS}$ curves of NMOS and PMOS devices at $T = 77$ K (blue circles) and 298 K (red squares) (a) before and (b) after the halo implant optimization.

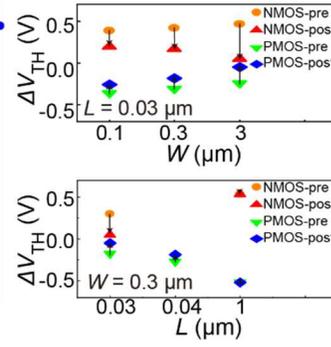

Fig. 5. The effect of halo implants on tuning $V_{TH}$ in devices with varied channel length and width at 300K.

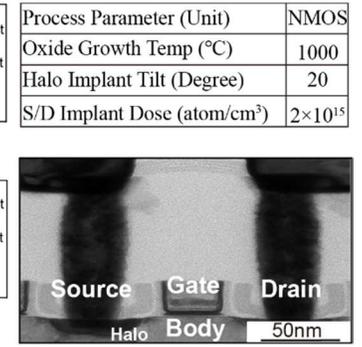

| Process Parameter (Unit) | NMOS |
|---|---|
| Oxide Growth Temp (°C) | 1000 |
| Halo Implant Tilt (Degree) | 20 |
| S/D Implant Dose (atom/cm³) | $2 \times 10^{15}$ |

Fig. 6. Key process parameters based on TCAD simulations and the TEM image of the fabricated Cryo-CMOS device.

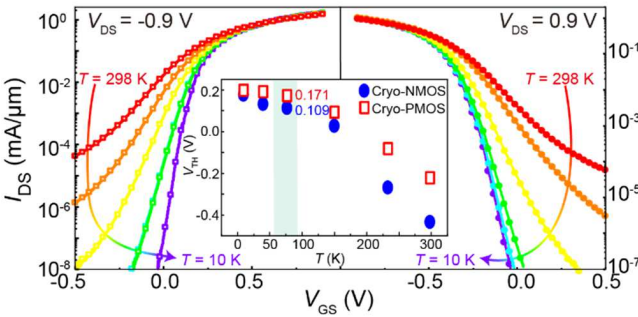

Fig. 7. Temperature-dependent $I_{DS}$-$V_{GS}$ transfer characteristics of optimized Cryo-NMOS and Cryo-PMOS devices from $T = 10$ K to 298 K. Inset: Temperature-dependent threshold voltage extracted by the constant current method in the saturation region ($V_{DS} = 0.9$ V).

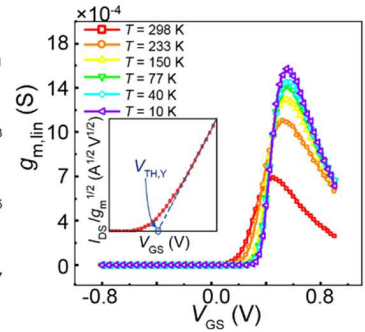

Fig. 8. Temperature-dependent $g_{m,lin}$ in the linear region ($V_{DS} = 0.05$ V). The $V_{TH,Y}$ value can be obtained by linear extrapolating the $I_{DS}/\sqrt{g_m}$ curve.

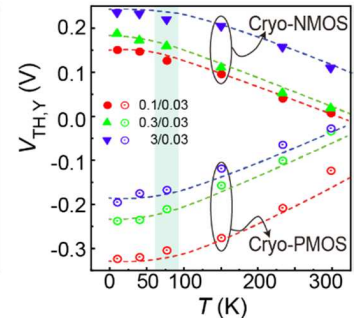

Fig. 9. Threshold voltage extracted from the Y-function method in Cryo-NMOS and Cryo-PMOS devices with varied channel width.

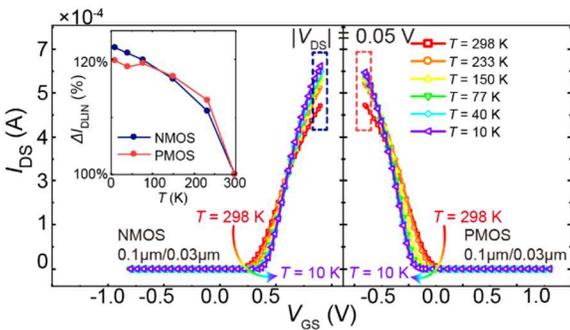

Fig. 10. Temperature-dependent $I_{DS}$-$V_{GS}$ curves of nanoscale Cryo-NMOS/PMOS devices at $V_{DS} = 0.05$ V. Inset: The evolution of the channel current as a function of temperature.

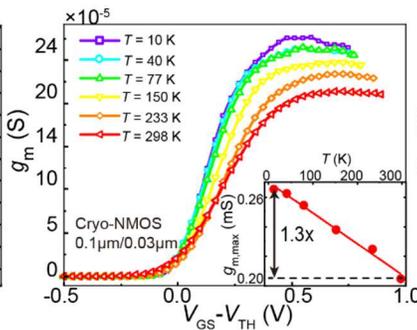

Fig. 11. Measured $g_m$ as a function of $V_{GS}$–$V_{TH}$ at different temperatures. Inset: The maximum transconductance increases by 30% at $T$=10 K.

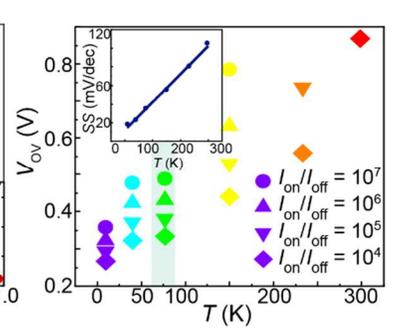

Fig. 12. The overdrive voltage corresponds to the on/off ratio from 10K to 298K. Inset: the $SS(T)$ curve of the Cryo-NMOS device.

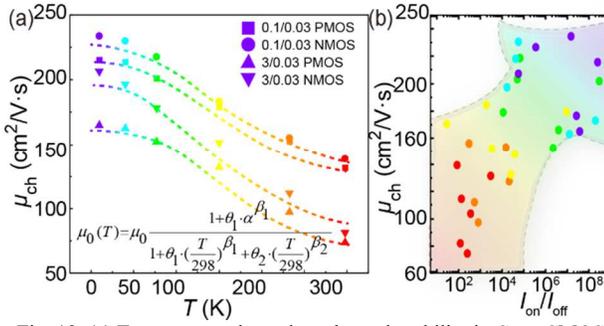
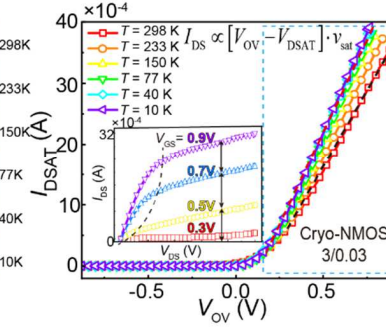
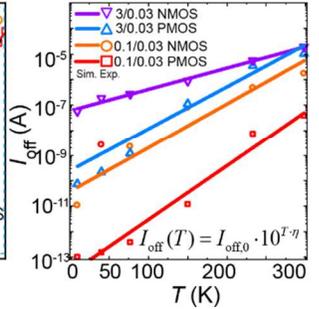

Fig. 13. (a) Temperature-dependent channel mobility in Cryo-CMOS devices. The saturation of $\mu_{ch}$ in the LT region is due to the Coulomb scattering. (b) Summary of $\mu_{ch}$ and $I_{on}/I_{off}$ dataset from 10K to 298K.

Fig. 14. Verification of the velocity saturation effect in the Cryo-CMOS device. Inset: Recorded $I_{DS}$-$V_{GS}$ data at 77 K.

Fig. 15. Universal scaling law of the leakage current with respect to temperature.

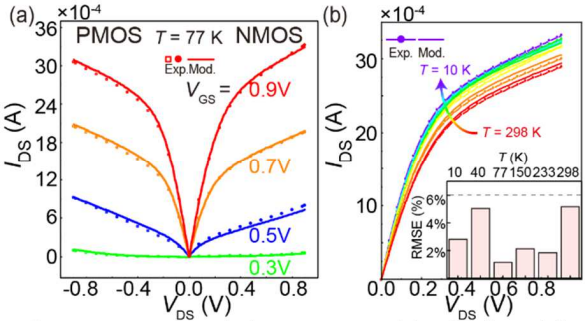
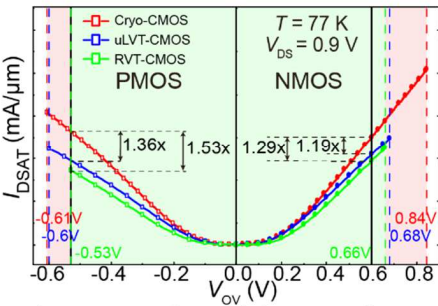
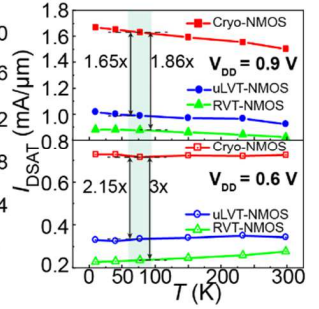

Fig. 16. Cryo-CMOS device compact model validation. Fitting results of (a) NMOS and PMOS at 77 K, and (b) $I_{DS}$-$V_{GS}$ at different temperatures. Inset: average model fitting error is well-below 6%.

Fig. 17. Comparisons of the measured $I_{DS}$-$V_{OV}$ curves between Cryo-CMOS, uLVT, and RVT-type transistors at $T$ = 77 K.

Fig. 18. Comparisons of saturated $I_{DSAT}$ values between NMOS/PMOS transistors at $V_{DD}$ = 0.6 V and 0.9 V.

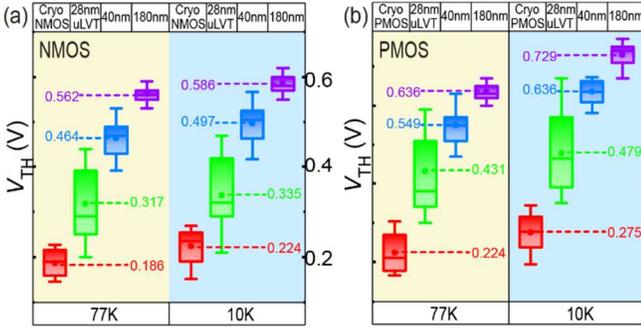
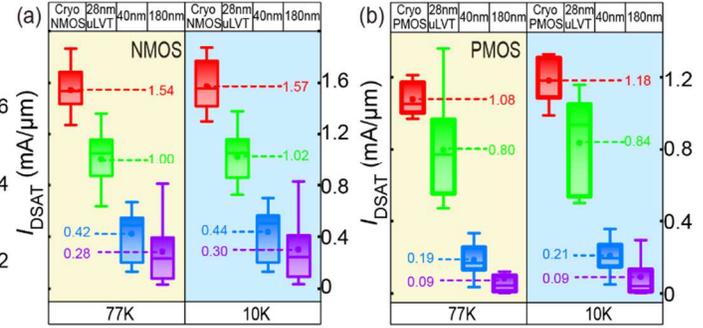

Fig. 19. Benchmarking the threshold voltage $V_{TH}$ among (a) NMOS and (b) PMOS devices from different CMOS process technologies and varied device sizes at $T$ = 10 K and 77 K.

Fig. 20. Benchmarking the saturation current $I_{DSAT}$ among (a) NMOS and (b) PMOS devices from different CMOS process technologies and varied device sizes at $T$ = 10 K and 77 K.

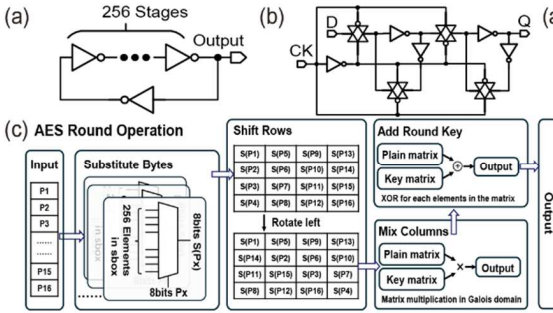
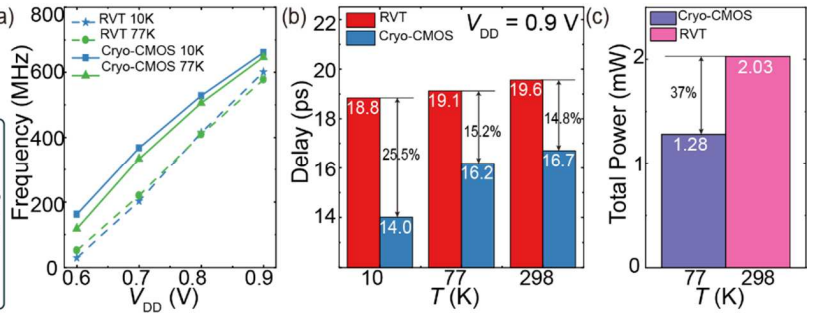

Fig. 21. Schematics of (a) ring oscillator, (b) D-flip flop, and (c) circuit block diagram to implement the AES algorithm.

Fig. 22. (a) Oscillation frequency of the cryogenic RO under different supply voltages. (b) Comparison of the DFF propagation delay realized by Cryo-CMOS and RVT devices. (c) Total power of the AES systems with the same frequency but different temperatures.